# In vivo Adaptive Focusing for Clinical Contrast-Enhanced Transcranial Ultrasound Imaging in Human

Justine Robin*[1,2], Charlie Demené*[1,2], Baptiste Heiles[1], Victor Blanvillain[1], Liene Puke[2], Fabienne Perren-Landis[†2], Mickael Tanter[†1]

[1] Physics for Medicine Paris, Inserm, ESPCI Paris, PSL Research University, CNRS

[2] Neurocenter of Geneva, LUNIC Laboratory, University of Geneva, Switzerland

*Contributed equally to this work

[†] Contributed equally to this work

**Email**:
 justine.robin@espci.fr

**Abstract**

Imaging the human brain vasculature with high spatial and temporal resolution remains challenging in the clinic today. Transcranial ultrasound is scarcely used for cerebrovascular imaging, due to low sensitivity and strong phase aberrations induced by the skull bone that only enable major brain vessel imaging, even with ultrasound contrast agent injection (microbubbles). Here, we propose an adaptive aberration correction technique for skull bone aberrations based on the backscattered signals coming from intravenously injected microbubbles. Our aberration correction technique was implemented to image brain vasculature in adult humans through temporal and occipital bone windows. For each patient, an effective speed of sound, as well as a phase aberration profile, were determined in several isoplanatic patches spread across the image. This information was then used in the beamforming process. It improved image quality both for ultrafast Doppler imaging and Ultrasound Localization Microscopy (ULM), especially in cases of thick bone windows. For ultrafast Doppler images, the contrast was increased by 4dB on average, and for ULM, the number of detected microbubble tracks was increased by 38%. This technique is thus promising for better diagnosis and follow-up of brain pathologies such as aneurysms or stroke and could make transcranial ultrasound imaging possible even in particularly difficult-to-image patients.

## 1. Introduction

Imaging the human brain vasculature with high enough spatial and temporal resolutions to detect small vessels and monitor their hemodynamics is still a challenge today in the clinic. Accessing this information would yet be of great interest for the diagnostic, follow-up, and better understanding of various cerebral pathologies such as aneurysms, or stroke-related events, but also in several degenerative pathologies such as Alzheimer's disease [1]–[3]. The gold standard imaging methods generally require contrast injections, and ionizing (CT) or expensive (MRI) imaging devices. They provide resolutions in the 0.4-0.34 mm range [4], [5], when the small vessel diameters are in the 10 μm range. Ultrasound is traditionally scarcely used in neuroimaging due to its very limited sensitivity and resolution at frequencies passing through the skull bone. Transcranial Color Doppler (TCCD) Ultrasound thus provides low-resolution images, where only the proximal segments of the cerebral basal arteries (circle of Willis and afferent arteries) can be depicted [6].

Increasing the sensitivity of ultrasound to blood flow has been an extensive field of research over the last 20 years [7], entailing ultrasonic probe developments, modification of the ultrasonic emission schemes [8], and refining of the signal processing chain [9]. An interesting setting for ultrasensitive blood flow imaging combines Ultrafast imaging and spatiotemporal Singular Value Decomposition (SVD) [10], with proven efficiency for cerebrovascular imaging [11].

More recently, ultrafast imaging combined with a contrast agent was proved able to beat by almost two orders of magnitude the resolution limit of conventional Ultrasound, a technique named Ultrasound Localization Microscopy (ULM) [12]–[15]. Inspired by super resolution optical microscopy techniques [16], the idea is to use microbubbles, a widely used ultrasound contrast agent, as strong point-like scatterers distributed in the vasculature [17], [18]. These microbubbles indeed have a typical 2-3 µm mean diameter but are yet very echogenic thanks to the large acoustic impedance mismatch between gas and liquid. Imaged by ultrasound at thousands of frames per second, they can thus be individually localized and followed in time from one frame to the next, providing vasculature maps resolving structures as small as 9 µm, highly surpassing the ultrasound diffraction limit.

If these methods have successfully been used in rodents, their translation to the clinic remains challenging, due to the much higher imaging depth required, and the huge obstacle of the skull bone. Bones indeed exhibit much higher density ($\rho_{bone} \sim 1000\ to\ 2200\ kg.m^{-3}$) and sound speed ($c_{bone} \sim 2700\ to\ 3000\ m.s^{-1}$) than soft tissues such as skin, muscle, or brain ($\rho_{soft} \sim 1000\ kg.m^{-3}$, $c_{soft} \sim 1480\ to\ 1540\ m.s^{-1}$), which leads to a high acoustic impedance mismatch and poor acoustic transmission at the skin/bone and bone/dura mater interfaces. Furthermore, the acoustic attenuation coefficient inside the skull bone is one of the highest in the human body (~30 dB/cm at 2 MHz), which means that imaging is only feasible through the thinnest point, the temporal window. Finally, the skull biases the image reconstruction by modifying the speed of sound on a section of the propagation medium, and the spatial heterogeneities of the skull bone width and sound speed lead to phase and amplitude distortions in the transmitted and received wavefronts, heavily affecting image quality. To overcome this problem, a large number of phase aberration correction techniques have been developed [19], [20]. They usually model the skull as a thin phase – or a phase and amplitude – screen located in the near field of the transducer. The relative temporal delays to apply to each transducer element are calculated to correct the aberrations. Most techniques obtain these delays using the correlation between signals received on different elements. They then iteratively repeat the process to increase the spatial coherence of the signals based on an indicator such as the focus criterion [21] or speckle brightness [22]. They can take advantage of a point-like scatterer if one is present in the medium [23], [24], or use diffuse scatterers if not. In particular, in the case of blood flow imaging, several methods using moving scatterers have been developed [25], [26]. Their implementation in the clinic however remains challenging, and in the case of transcranial Doppler, is still limited to the major brain vessels [27], [28].

In the particular case of transcranial ULM however, the microbubbles not only compensate for the high attenuation of the skull bone but also enable the correction of the strong skull aberrations. Each micro-bubble can indeed be considered as a point-like source conveniently placed behind the skull. This microbubble can be used as a beacon for the recovery of the skull bone aberration by providing an experimental estimation of the Green's function relating the microbubble position to the piezoelectric elements of the ultrasonic array. The skull-induced phase aberration profile, as well as the effective speed of sound of the medium, can thus be derived by studying the distortions in the wavefront coming from this beacon. This technique, in a very preliminary form, was briefly mentioned in the first proof of concept of transcranial ULM in human [15]. In the present paper, we take this technique to an accomplished level, giving extensive details about its implementation, showing results both on contrast transcranial Doppler and transcranial ULM images, and quantifying the improvement on the images, for application in cerebro-vascular imaging in human adults.

## 2. Materials and Methods

### 2..1 Clinical Protocol

All experiments strictly comply with the ethical principles for medical research involving human subjects of the World Medical Association Declaration of Helsinki. Healthy volunteers and patients were recruited under the protocol accepted by the CCER of Geneva (n° 2017-00353) and gave informed and written consent. The dose of injected contrast agent as well as the amplitude and duration of ultrasound exposures were kept to the minimum enabling the ultrasound localization microscopy to follow the ALARA ("As Low As Reasonably Achievable") principle. Ultrasound parameters were well below the FDA recommendations (AUIM/NEMA 2004, Track 3) for ultrasound imaging, with a maximum Mechanical Index (MI) of 0.46 (maximum FDA recommended value is 1.9), a maximum derated Spatial Peak Temporal Average Intensity (ISPTA) of 64.3mW/cm² (maximum FDA recommended value 720mW/cm²), a maximum Thermal Cranial Index (TIC) of 1.99 (FDA regulations ask for explanations for values above 6). A widely clinically used echo-contrast agent consisting of Sulphur hexafluoride micro-bubbles with a mean diameter of 2.5 µm and a mean terminal half-life of 12 min (SonoVue®, Bracco, Italy) was injected intravenously via the cubital vein. Up to three 0.1 mL boli of contrast agent were injected successively, corresponding to a maximum bubble concentration of 6 to $50.10^6 \, microbubble/L$ in the blood. The brain vasculature of adult healthy volunteers (27 to 82 years old, age median: 79) was imaged transcranially through the temporal or occipital bone windows.

### 2.2 Ultrasound acquisitions

A phased array ultrasonic probe (XP 5-1, pitch 0.2 mm, 96 elements, central frequency 2.93 MHz, 90 % bandwidth at -6dB) (Vermon, Tours, France) and an ultrafast programmable ultrasound scanner (Aixplorer® Supersonic Imagine, Aix-en-Provence, France) were used for ultrafast ultrasound imaging. The imaging sequence consisted of 4 successive diverging waves originating from different virtual sources (regularly spaced every 3.2 mm, placed 11.44 mm behind the transducer), emitted at a frequency of 2 MHz and a PRF of 4800 Hz (compound framerate of 800 Hz). For each emission, backscattered echoes were recorded by the transducer array, digitized at 200% bandwidth (meaning 4 samples per wavelength), and stored in a so-called radiofrequency (RF) data matrix. A 1 s emission was repeated every 2 s, for a total acquisition time of 90 s.

### 2.3 Image reconstruction and bubble localization

The images were reconstructed using delay-and-sum beamforming with and without integration of the calculated phase aberration law. The 3D matrix ($space \times space \times time$) of the full stack of images was then filtered to extract signals coming from the microbubbles. First, an SVD clutter filtering was used to remove tissue signals [29]. Depending on the level of tissue motion, between 25 and 50 singular vectors out of 800 were removed following the method described in [10]. Then, a binary mask was built based on the vesselness filtering of this stack of images [30] (available on Mathworks file exchange, ©Dirk-Jan Kroon 2009, and © Tim Jerman, 2017). In the ($space \times space \times time$) 3D matrix, moving bubbles will indeed appear as tubular (vessel-like) structures and will therefore be enhanced. Both image stack and mask stack were interpolated (Fourier Space based interpolation (equivalent to an exact sinc interpolation) for the image stack, nearest-neighbor interpolation for the binary mask stack (the goal of this interpolation being keeping consistent dimensions for the image stack and the binary mask stack) to obtain a radial resolution of $dR = \lambda/6$ and an angular resolution of $d\varphi = 0.5°$. Local maxima were then detected within the masked area for each frame. Small regions around these local maxima were then correlated with the typical point spread function of our imaging system (which is the response of an isolated microbubble), and only maxima with strong correlation (>0.6) were kept. Localization was further refined at the sub-pixel level using a fast local (5x5 pixel neighborhood) 2nd order polynomial fit, and their positions were finally converted from polar to Cartesian coordinates. The maxima positions were then tracked using a classical particle tracking algorithm (simpletracker.m available on Mathworks ©Jean-Yves Tinevez, 2019, wrapping Matlab munkres algorithm implementation of ©Yi Cao 2009) with no gap filling and maximal distance linking of 1 mm (corresponding to a bubble maximum speed of 80 cm/s). To reduce further the level of false bubble detection, bubble tracks shorter than 10 frames were removed, based on the idea that microbubbles traveling in the bloodstream should be observed on several consecutive frames [31].

### 2.4 Aberration correction method

Using isolated bubbles as individual ultrasound point-like sources placed directly inside the patient's brain vasculature, an iterative aberration correction procedure was developed to both estimate the effective sound speed of the imaged medium (average over skull and brain tissues in the propagation path) and to determine the phase aberration profiles introduced by the skull. Our first goal was to validate (or invalidate) the hypothesis modeling the skull as a thin phase screen aberrator – in which case the aberration law can be considered the same over the whole image. Phase aberration laws were thus first calculated using bubbles distributed over the whole image, to identify potential isoplanatic patches within which the phase aberration remained constant (within a patch, the aberration law variation across bubbles is less than T/8, T being the period of the ultrasound wave). In each of these patches, an iterative procedure was then used, where phase aberrations and effective sound speed were calculated in turn, until convergence of the effective sound speed value (see Fig. 1).

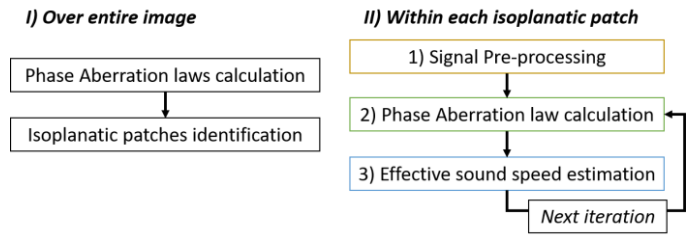

Fig. 1. Description of the aberration correction process. After identification of the different isoplanatic patches, an effective sound speed and a phase aberration law are recovered in each of them.

*Signal pre-processing*

SVD clutter filtering was performed on the raw RF signals, to remove tissue signals. Isolated bubble wavefronts appeared visible in the filtered RF data (see Fig 2, step 1). Filtered RF data were beamformed (spherical delay laws, $sound \, speed = 1540 \, m.s^{-1}$) and isolated bubbles were located in the uncorrected image (as described above). Bubble locations - $(R, \phi)$ in polar coordinates - were stored and used in the aberration correction procedure.

*Phase aberration law calculation*

To limit the influence of potentially overlapping wavefronts, only isolated microbubbles were used in the following algorithm. Namely, a bubble should be at least 2mm away from any other strong scatterer to be included. For each considered bubble, the procedure can be decomposed in 3 steps, and repeated in 5 iterations (see Fig. 2):

1) The filtered RF signals from the 4 different transmission events were delayed by time $r_s/c_0$, where $r_s$ is the distance between the bubble location and the virtual source s used to create each diverging wave emission. This creates a virtual emission focusing directly on the bubble. A directional filter is applied to the virtually focused RF signals - in k-space – to isolate the wavefronts specifically coming from the bubble location from potential overlapping wavefronts. This is done by computing the 2D Fourier transform of the compounded RF data, which gives a k-$\omega$ diagram, k being the wave vector and $\omega$ the angular frequency, and by setting to 0 a certain range of orientations before computing the invers 2D Fourier transform. This will select the waves propagating in a certain direction (with a certain k vector).
2) For each receiving sensor i signal is shifted by the delay $r_i/c_0 + \tau_{i,n}$, where $\tau_{i,n}$ is the correction for iteration n. The phase aberration profile is determined by finding the maximum cross-correlation of the signals between transducer elements (see Section 2-5).
3) These delays are added to the focusing law used in step 1, for the next iteration.

At each iteration, the spatial coherence function is calculated in the filtered RF at the microbubble position, to evaluate aberration correction performance (see Section 2.5 for details). Bubbles for which the integral of the spatial coherence does not increase are discarded for the rest of the process. An average phase aberration law is calculated for each isoplanatic patch and used for the next sound speed estimation step. Finally, at the end of the process, the strength of aberration is evaluated for each acquisition as the root-mean-square (RMS) of the calculated aberration profile. The whole method is detailed in Fig. 2.

*Effective sound speed estimation*

In this step, different effective sound speeds varying from $c_i = 1450\ m.s^{-1}$ to $c_i = 1650\ m.s^{-1}$ are screened. For each sound speed considered:

1) The bubble positions - $(R, \phi)$ in polar coordinates - obtained from the uncorrected beamformed images are corrected to account for the tested sound speed $R_i = R \cdot \frac{c_i}{1540}$, $\phi_i = \phi$.
2) The filtered RF signals from the 4 different diverging wave emissions are recombined using the delay laws corresponding to bubble location at the considered sound speed, including the phase aberration calculated in a previous step. A directional filter is applied to the focused RF signals – in the k-space – to isolate the wavefronts specifically coming from the bubble location from potential overlapping wavefronts.
3) The spatial coherence function is calculated at the bubble location and stored.

The spatial coherence functions obtained for each tested sound speed are averaged over all the considered bubbles, and the effective sound speed exhibiting the highest area under the spatial coherence function curve is picked and used in the next iteration. The algorithm stops when the same sound speed is obtained in 2 consecutive iterations.

*Corrected image reconstruction*

At the end of the aberration correction protocol described above, we have identified for each isoplanatic patch of the image a couple (effective sound speed, phase aberration law). This information is thus used to beamform as many corrected images as isoplanatic patches. A final composite corrected image is then reconstructed by combining all the corrected patches.

## 2.5 Aberration correction evaluation

The robustness of our aberration correction method was evaluated in several manners, to ensure that the calculated phase aberration laws were meaningful. First, the convergence of the algorithm was tested in terms of the number of iterations needed to converge towards the final delay law. To do so, the performance of the correction algorithm with an increasing number of iterations was quantified directly on the RF signals.

For each considered bubble, the spatial coherence function $R(m)$ - defined by Van Cittert and Zernike as the average cross-correlation between signals received at two points of space (here the transducer elements positions) [32] – was calculated on the filtered and flattened RF data [33], [34]:

$$R(m) = \frac{N}{N-m} \frac{\sum_{i=1}^{N-m} u(i, i+m)}{\sum_{i=1}^{N} u(i, i)}$$

where m is the distance in transducer elements, N is the number of elements in the transducer, and $u(i,j)$ is defined as:

$$u(i,j) = \sum_{T_1}^{T_2}(S_i(t) - \overline{S_i})(S_j(t) - \overline{S_j}),$$

with [$T_1$, $T_2$] a temporal window centered on the focal time, and $S_i$ is the time-delayed (flattened by subtraction of the parabolic time delay law in the homogeneous medium corresponding to the effective sound speed) RF signal received on transducer $i$.

The Van Cittert Zernike theorem indeed states that the coherence function is the spatial Fourier transform of the intensity distribution at the focus. If the emission signals are recombined using the correct focusing law, and if the RF signals are then flattened with the correct delay law, the microbubble behaves as a point-like target, backscattering a coherent wavefront. The coherence function should thus be a constant step function across the transducer aperture [21].

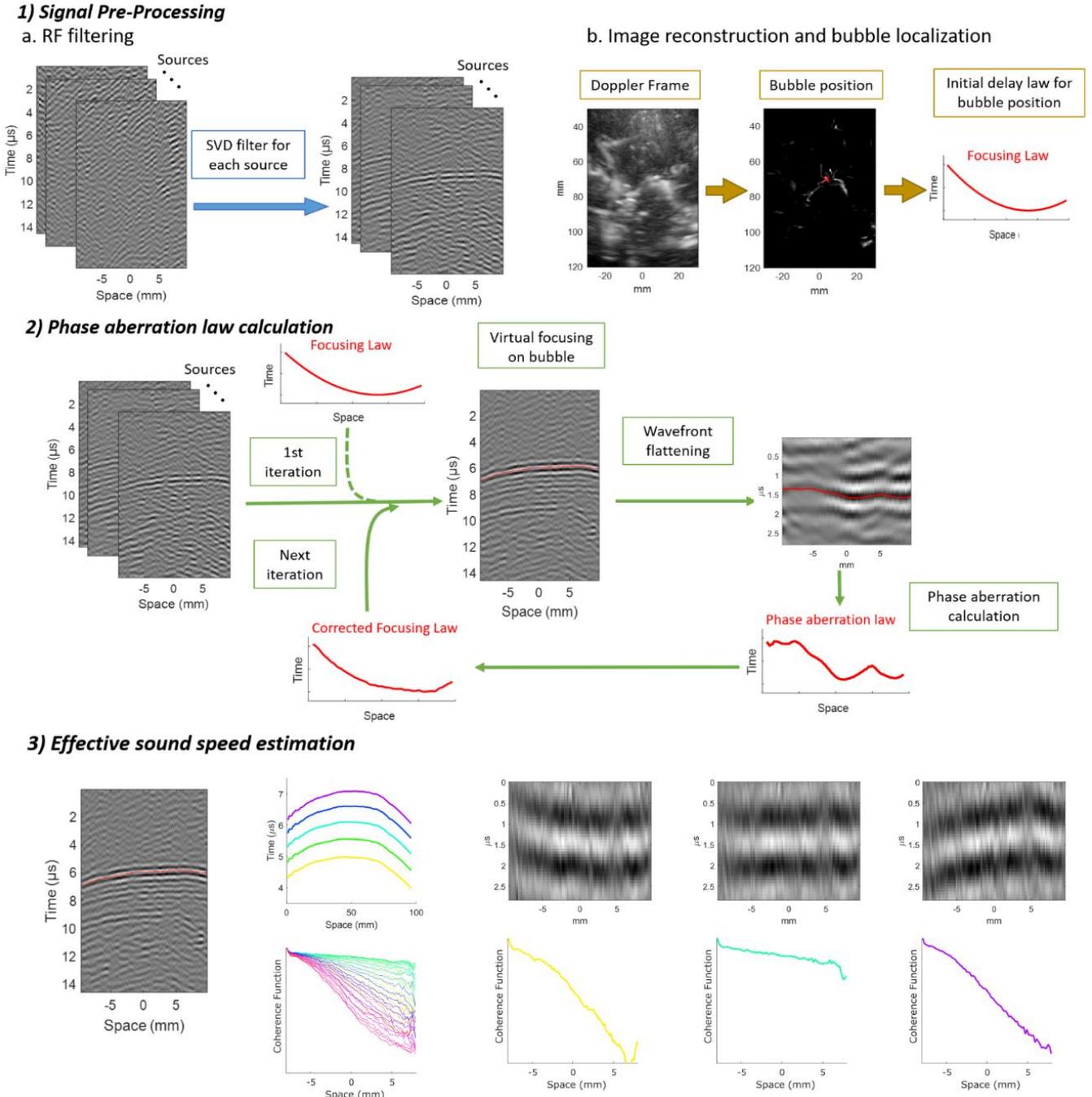

Fig. 2. Algorithm workflow in each isoplanatic patch. 1) Signal pre-processing: SVD clutter filter is first applied to the raw RF data to remove tissue signal and let the microbubble wavefronts appear. Power Doppler images are reconstructed from the filtered RF data, and bubble positions (red cross) and the corresponding spherical delay laws are calculated and stored to initialize the iterative algorithm. 2) The phase aberration law is calculated for each bubble: the last calculated delay law is used to virtually focus on the bubble position, and to flatten the wavefront originating from the bubble. The phase aberration law is obtained from the flattened wavefronts using cross correlation. This aberration law is used to correct the spherical delay law for the next iteration. 3) Effective sound speed estimation: wavefronts originating from each bubble are flattened using delay laws calculated with different sound speeds (1450 to 1650 m/s) and the last calculated aberration law. The coherence function of these wavefronts are then computed, and the one exhibiting the highest coherence is selected as the effective sound speed and used for the next iteration.

More generally, the area under the coherence function curve increases as the phase aberration delays are corrected. This area was thus calculated at each iteration of the algorithm to determine the optimal number of iterations needed, selected as the one after which the area stopped increasing.

As mentioned above, the phase aberration laws were calculated on average over several bubble localization events within the same isoplanatic patch. The aberration laws were thus computed using an increasing number of localization events, to estimate how many bubbles should be used on average. In each case, the mean coherence function was calculated on the bubble wavefronts, to assess the performance of the correction. The root mean square (rms) of the difference between these aberration laws was also computed, as a marker of the convergence of the method.

To further confirm this convergence towards a meaningful phase aberration law, two acquisitions were made in the same imaging plane, on the same patient, but rotating the ultrasound probe by 180°. After flipping the 2$^{nd}$ acquisition, the obtained aberration laws are thus expected to be similar. This similarity was evaluated as the rms of the difference of the 2 aberration laws.

Finally, the image quality improvement obtained with our aberration correction method was evaluated both qualitatively and quantitatively on the PW Doppler image and the localization image.

*On the beamformed image*

The image quality improvement was assessed on the power Doppler images qualitatively by observing the delimitation of small vessels in the near and far fields, and quantitatively by evaluating the contrast to noise ratio: $CNR = \frac{|S(ROI_1) - S(ROI_2)|}{std(ROI_2)}$, where $ROI_1$ is a region of interest defined in a large vessel on the PW Doppler image, and $ROI_2$ is a region of interest defined in the background noise next to this vessel.

*On the localization microvascular image*

Once the images are reconstructed after aberration correction, bubble localization and tracking is performed again on those corrected images as explained in section 2.3. The full localization image is built by tracking all the localized bubbles across the frames. Then bubbles with tracks shorter than a given threshold length (10 frames for instance) are discarded, based on the idea that microbubbles should be observed on several consecutive frames, and only the longer tracks are kept and aggregated to make the final image. The more bubbles we can find, and the longer their tracks are, the better the image will be, as more information on the position and hemodynamics of the vessels will be collected. We thus compared the number of bubbles found in the corrected and uncorrected images, as well as their track lengths. Namely, bubbles exhibiting a track longer than different thresholds (10, 20, and 30) were counted in both cases.

### 3) Results

*3.1) Algorithm convergence speed and robustness*

An example of aberration law obtained in a representative localization event with an increasing number of iterations is displayed in Fig. 3, along with the corresponding wavefront coherence functions. We observe that the area under the coherence curve (Fig. 3c) increases with the number of repetitions. After 5 iterations, it reaches a plateau, and the coherence function approaches the step function corresponding to the un-aberrated case (Fig. 3b). This number of iterations was thus consistently used for the rest of the study.

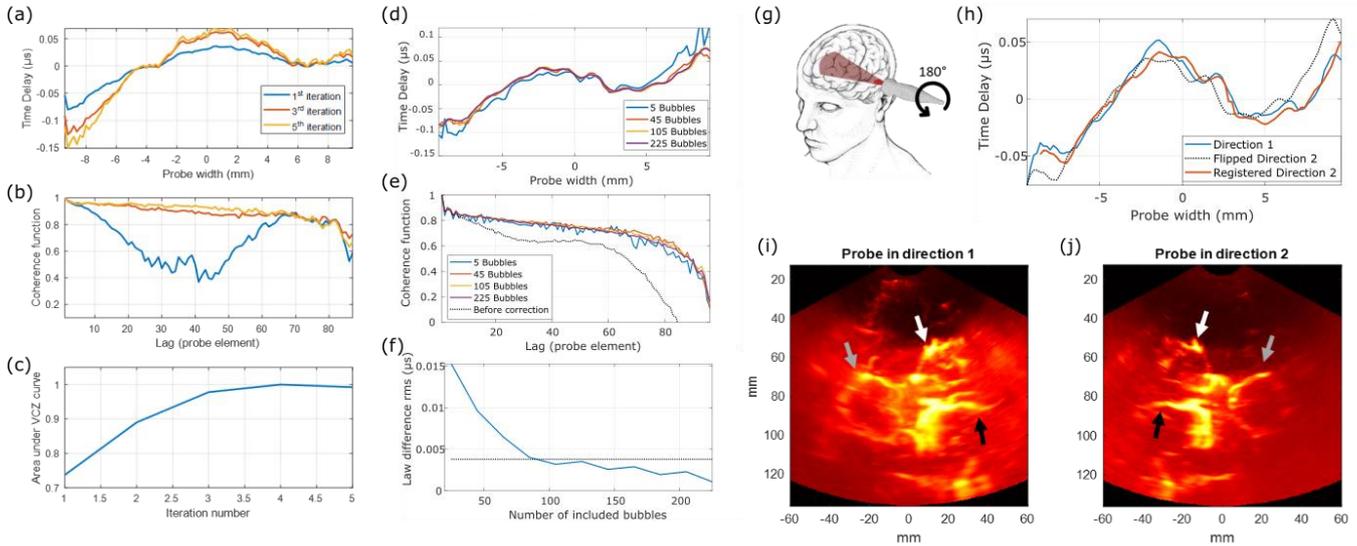

Fig. 3. Algorithm assessment. Left: Influence of the number of iterations. Example of (a) phase aberration laws obtained after 1, 3 and 5 iterations along with (b) the corresponding coherence function calculated on the transducer elements, and (c) the evolution of the area under the coherence function across the iterations. Center: Influence of the number of bubbles. (d) Example of phase aberration laws obtained with 5, 45 and 105 and 225 microbubbles from the same isoplanatic patch. (e) Corresponding coherence function calculated on the transducer elements. The uncorrected coherence function is plotted for comparison (f) Rms of the laws difference when including 20 additional bubbles. The dashed black line represents 10% of the final aberration law rms. Right: Validation on a healthy volunteer. (g) Drawing schematizing the acquisition process, and showing the left trans-temporal insonation plane in a healthy adult volunteer. The ultrasound probe is applied on the temporal bone in the conventional direction of an axial plane (image (i)) and then turned by 180° (image (j)). (h) Aberration laws obtained in the 2 directions. Unconventional direction is flipped for easier comparison, and the rigid transformation between the 2 images is applied. (i) Power Doppler image obtained with the probe in the conventional direction and (j) Power Doppler image obtained with the probe flipped by 180°. The arrows show mirror structures: white arrow for the medium cerebral artery, gray full arrow for the anterior cerebral artery, and black arrow for the posterior cerebral artery.

The number of bubbles needed to converge to a stable phase aberration law was evaluated next. For each acquisition, an increasing number of bubbles were selected in a given isoplanatic patch and the corresponding average aberration laws were calculated. The average coherence function obtained after correction with these different laws was also computed. A representative example is displayed in Fig. 3 d and e. With as few as 5 bubbles, the coherence function already approaches the step function, and the aberration law doesn't change much above 50 to 100 included bubbles. To quantify this, Fig. 3f shows the rms of the aberration law difference when including 20 additional bubbles, we observe that above 85 bubbles, this difference is below 10% of the total law rms (black dashed line). On average over all acquisitions, 88 ±19 bubbles are needed to reach this threshold.

Fig. 3 g to j shows the results of the aberration correction process on the same patient (27 years old), in the same imaging plane, with the ultrasonic probe in the conventional direction and rotated by 180°. The 2 images were registered (rigid transformation), and the transformation was then applied to their respective phase aberration laws. As expected, these laws are extremely similar, with rms difference of 6 ns - 18% of the total law rms - which can be explained by a slight difference between the two imaging planes. This similarity emphasizes the robustness of the proposed method.

### 3.2) Isoplanatic patches determination

Phase aberration laws calculated on bubbles distributed over the whole image were compared along the radial distance to the ultrasonic probe (R) at a fixed angle, and for varying angle positions ($\varphi$) at a fixed radial distance. In both cases, patches of similar sizes were defined, and the average aberration law across all bubbles in the patch was calculated. The level of similarity of these average laws was evaluated by computing the rms of their differences and their correlation. It is usually considered that phase variations below T/8 do not affect image quality [35], and this criterion was thus used to separate the different isoplanatic patches. For all patients, the phase aberration law appeared to strongly depend on $\varphi$ only. An example acquisition is displayed in Fig. 4, along with the corresponding aberration laws, and their correlation as a function of the angular or radial distance between patches. This result tends to prove that our initial hypothesis modeling the skull as a thin screen phase aberrator is only partially appropriate, as the aberration law is consistent across depths for a given angle. Considering the skull as a thick aberrator is however more exact, as the angle between the incoming wave and the skull influences the aberration law.

In the rest of the study, we thus identified isoplanatic patches where the aberration law was considered constant as angular sectors covering the whole image. With this hypothesis, between 1 and 4 non-overlapping isoplanatic patches were identified for each patient. In more detail, 2.1±1 patches were identified on average over the 10 acquisitions included in the study, a number that goes up to 2.5±1 patches when considering only transtemporal axial views, and down to 1.5±0.5 patches when considering only transtemporal coronal and occipital views.

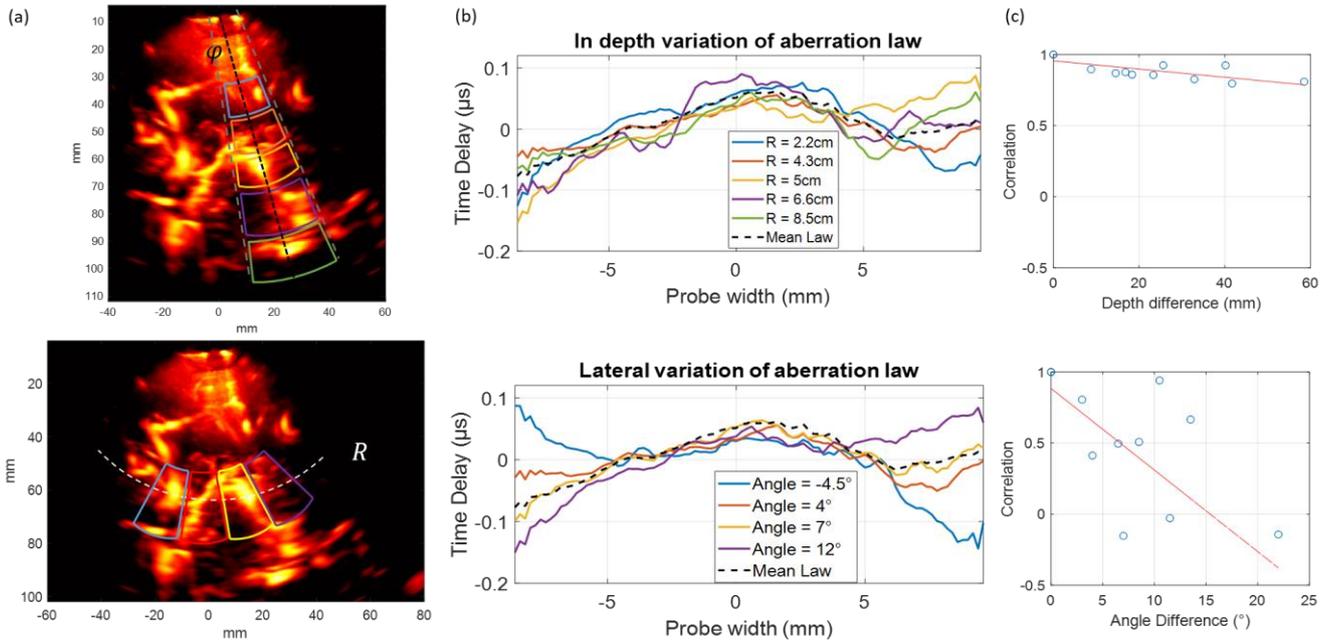

Fig. 4. (a) Example of a trans-temporal power Doppler image showing the different patches in which phase aberration laws were compared. (b) Mean aberration laws calculated in each patch. (c) Correlation between the mean aberration laws as a function of the angular or radial distance between the different patches.

*3.3) In-patch sound speed estimation*

In each of the isoplanatic regions defined in the previous step, our iterative method was implemented, to determine the couple (effective sound speed, phase aberration law) providing the highest spatial coherence in the bubble wavefronts. For the same patient, Fig. 5 displays the spatial coherence obtained for a range of sound speeds in each of the isoplanatic regions without phase aberration correction and after the last iteration of our algorithm. Before correction, it is very difficult to even identify a relevant effective sound speed as the coherence is strongly affected by skull aberrations. After correction, the spatial coherence is globally increased by 27±8 % on average over all acquisitions (see Table 1 for detailed results), and a coherence maximum can be identified, to choose the appropriate effective sound speed to use in the beamforming process. This sound speed value (1550 m/s in the displayed example) integrates the contributions of all the tissues in the wave propagation path: the influence of both the skull with high sound speed (up to ~ 3000m/s) and the brain tissues with lower sound speed (~ 1520 m/s [36]) are combined.

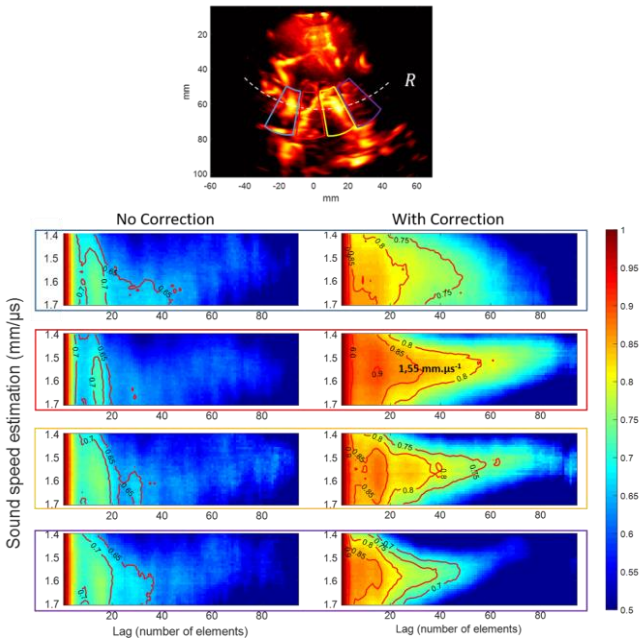

Fig. 5. Spatial coherence of the bubble wavefronts obtained in each isoplanatic patches before and after implementation of the iterative aberration correction algorithm performed using 5 iterations, and 150 microbubbles. The same colormap is used in both images.

*3.4) Overall image quality improvement*

Fig. 6 shows an example of a Power Doppler trans-temporal image in axial view with and without implementation of our aberration correction protocol. Qualitatively, vessels are better defined at different depths in the image. The near field that is usually quite blurry is sharpened: a descending vessel appears thinner, while the noise level in the background is reduced (Fig. 6 – b). At a distance (65 mm, which corresponds to the elevation focal distance of the probe) sharpening of vessels is irrefutable, with even the disambiguation of vessels that were mixed in the case without correction (see around the coordinates (10, 65) mm in Fig. 6 – c). Quantitatively, for all acquisitions the spatial coherence was on average largely increased (from +14% to +40%) by our technique, meaning that we take greater advantage of the antenna gain of our transducer array. The CNR was improved for all acquisitions through temporal windows. It was increased by 4 dB on average on all the acquisitions, reaching a value of +9 dB in one example.

To assess the improvement in the localization images, the number of bubbles successfully localized and tracked across several frames were compared with and without correction.

After correction and on average over all acquisitions, the number of bubbles tracked across at least 10 frames is increased by 11.5 %, by 30.5 % for tracks longer than 20 frames, and by 38% for tracks longer than 30 frames. Supplemental Fig. 1 illustrates how the detected bubbles have much longer tracks with aberration correction, especially at high depth where the SNR is low. We hypothesize that this trend (higher improvement for the longest tracks) is because those long tracks are probably subdivided into shorter tracks in the absence of aberration correction. In isoplanatic patches with stronger aberration, tracks can indeed be interrupted because of intensity (or shape) loss of microbubbles. This result is very interesting for ultrasound localization microscopy images as the longest tracks are the most reliable to delineate small vessels. Results for all the acquisitions are summarized in Table 1.

The inter-subject variability of the retrieved aberration laws, represented by their root-mean-square (RMS), is also visible in the first line of Table 1. The thickness and shape of the bone windows indeed vary from patient to patient, but also depending on the position of the ultrasonic probe on the bone window (on the temporal window, the neurologist can for instance choose an axial or a coronal view), and the images can thus be more or less aberrated. Fig. 6 – d displays the bubble number increase as a function of phase aberration strength for the ten acquisitions. It is interesting to note that the stronger the aberration is, the more the final image is improved. A linear fit indeed gives a $1.84\%.ns^{-1}$ slope, with $R^2 = 0.8$.

Fig. 7 shows 3 examples of microbubble localization images for one trans-temporal axial (acquisition 8 in Table 1,

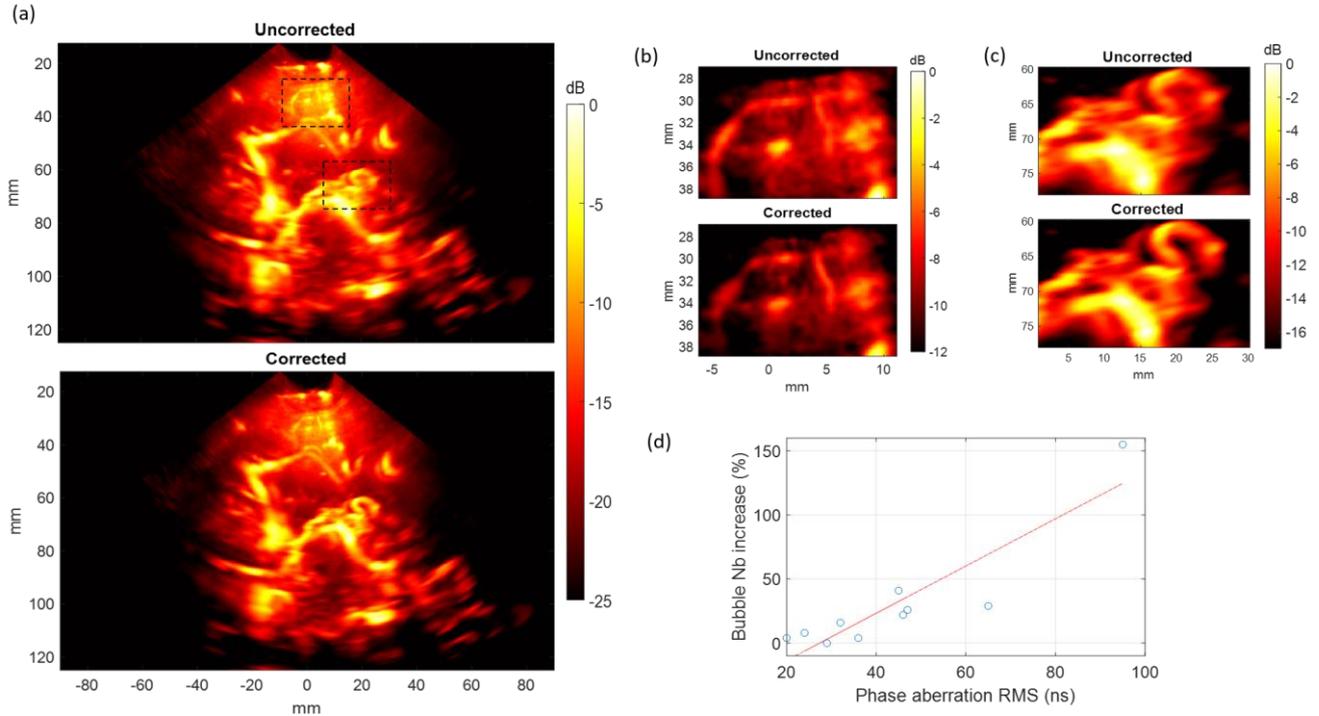

Fig. 6. Example of transcranial Power Doppler images obtained by ultrafast Doppler through the temporal bone with and without aberration correction. It corresponds to acquisition 4 in Table 2. (a) Full field of view, (b) zoom in the near field, (c), zoom at a higher depth; (d) Increase in the number of bubbles with tracks longer than 20 frames successfully localized for each acquisition, as a function of the phase aberration root-mean-square. Linear fit $y = a.x + b$ with $a = 1.84$, $b = -50$, and $R^2 = 0.8$

Fig.7 a, suppl. Fig. 2), and two trans-temporal coronal (acquisition 10 and 7 in Table 1, Fig. 7d and 7g respectively) views with and without aberration correction. The bubbles with tracks longer than 10 frames are represented. The first effect of aberration correction is to reveal previously invisible vessels: in Fig. 7b some small (~100 µm caliber) penetrating vessels become visible in the mesencephalon, and in Fig. 7c, vessels paths that appeared incomplete without aberration correction are improved by the technique. The second important effect of aberration correction is to strengthen some large vessels, especially when they are slightly out-of-plane or at depth (Fig. 7d, arrows). Indeed as the antenna gain is increased by the technique, the faint echoes of slightly out-of-plane bubbles are reinforced after correction and become detectable. A third important effect is the disappearance of "ghost" vessels, that sometimes appear as lateral replicates of large vessels in the uncorrected images (Fig. 7e and 7g). That can happen when part of the wavefront is so distorted that it contributes to a ghost echo with a wrong localization after beamforming. After correction, those echoes are removed, and therefore "ghost" vessels are suppressed. Supplemental figure 2 shows the MRA image of the patient corresponding to Fig. 7g – i, in the ultrasound imaging plane. It confirms the absence of the ghost vessel visible in Fig. 7g. Finally, for vessels of intermediate size (i.e. neither delineated by only a few bubbles nor by hundreds of them), the result of aberration correction is generally a sharpening of the structures (see Fig. 7f and 7i). Axial views generally give richer microbubble localization images, both because the temporal bone window is thinner in this orientation (meaning less severe aberration and attenuation) and because it gives access to many big vessels and their ramifications in the same plane. Therefore the main effect of aberration correction is the appearance of new vessels and the completion of discontinuities and sometimes strengthening of out-of-plane vessels. In coronal views, the aberration is generally stronger, and aberration correction can improve the image on the three aspects mentioned above.

### 4) Discussion

By taking advantage of microbubbles, injected as a contrast agent, we proposed a method for the estimation and correction of aberrations induced by the skull bone in ultrasonic data. The efficacy of the microbubble-based aberration correction was demonstrated in human adults using transcranial ultrafast ultrasound. The microbubbles were used as strong and isolated point-like scatterers conveniently placed behind the skull and distributed over the whole image. By means of these scatterers, the phase aberration profiles and effective sound speed variations introduced by wave propagation through the skull were determined.

These phase aberrations profiles were compared across the image and were found to strongly depend on the sectorial positions, but not on the imaging depth. These first findings partially invalidate the common hypothesis considering the skull as an infinitely thin screen phase aberrator, a hypothesis under which the aberration is the same in every part of the

TABLE I
QUANTIFICATION OF IMAGE QUALITY IMPROVEMENT AFTER ABERRATION CORRECTION

| Acq # | | | 1 | 2 | 3 | 4 | 5 | 6 | 7 | 8 | 9 | 10 |
|---|---|---|---|---|---|---|---|---|---|---|---|---|
| Age | | | 76 | 79 | 79 | 79 | 79 | 82 | 56 | 79 | 46 | 46 |
| View | | | Occip. | Temp. axial | Temp. axial | Temp. axial | Temp. axial | Temp. axial | Temp. coronal | Temp. axial | Temp. coronal | Temp. coronal |
| Phase aberration law RMS (ns) | | | 20 | 24 | 29 | 32 | 36 | 45 | 46 | 47 | 65 | 95 |
| Spatial coherence increase (%) | | | 27 | 20 | 27 | 21 | 26 | 40 | 14 | 35 | 38 | 20 |
| CNR increase (dB) | | | 0 | 1.2 | 4.6 | 5 | 2.3 | 5 | 9 | 3.4 | 4.8 | 5.3 |
| Bubble nb increase (%) | Min track length | 10 | 1.3 | 1 | -4 | 5.5 | 0 | 13 | 9 | 11 | 12 | 66 |
| | | 20 | 4 | 8 | 0 | 16 | 4 | 41 | 22 | 26 | 29 | 155 |
| | | 30 | 10 | 13 | 10 | 21 | 15 | 24 | 43 | 42 | 32 | 173 |

*Abbreviations: Occip. = occipital view; Temp. axial = transtemporal axial view; Temp. coronal = transtemporal coronal view*

medium. In consequence, for the rest of the study, several sectorial isoplanatic patches were identified in the images, and the aberration correction procedure was done independently within each patch.

The number and size of isoplanatic patches highly depend on patients, as well as cross-sectional views of the brain (i.e. transtemporal axial or coronal views and occipital view) as the quality and homogeneity of the acoustic window varies a lot from patient to patient and on a given patient depends on the probe positioning. The distribution of vessels over the field of view is also of great influence. A limitation of the proposed method is indeed that we can only compute the aberration delay laws in regions where a sufficient number of bubbles are located (in the order of 50 to 100 bubbles). In coronal or occipital views, for instance, vessels are typically mostly located in the center of the image (Fig. 7 (d) and (g)), making it more difficult to calculate meaningful aberration laws on the far sides of the field of view. For this reason, such cases typically feature 1 or 2 isoplanatic patches. Axial views on the other hand typically show vessels distributed on the entire field of view (Fig. 6 and Fig. 7 (a)), by which up to 3 or 4 patches can be defined.

Our aberration correction method enabled the improvement of:
1. The spatial coherence of the wavefronts in the RF data;
2. The contrast to noise ratio in the PW Doppler images;
3. The number of bubbles and their track lengths in the ULM microvascular maps.

It is interesting to note the high inter-subject variability of these improvements, and how the number of bubbles retrieved after correction correlates well with the RMS of the phase aberration laws. Skull bone characteristics such as density or thickness are indeed known to vary from subject to subject, and all cranial bone windows are not equivalent. Moreover, the natural bone windows are usually quite small in size, and the clinician sometimes has to place the probe very near their limits to access different parts of the brain, implying larger distortions of the wavefront. This seems to be particularly true for coronal views through the temporal window for instance. Here, the worse the aberration is, the more our technique improves the images, giving similar final image qualities in terms of penetration depth for instance. This result is thus quite encouraging for patients with poor bone windows that are today almost impossible to image.

Retrieving as many bubbles as possible will also be of particular importance for the mapping of the brain vasculature down to very small vessels. To image the microvasculature with ULM at clinically allowed microbubble concentrations, one important issue is indeed to ensure that a sufficient number of bubbles circulate all the way to the smaller vessels within a short acquisition time [37]. In this context, only a very small number of bubbles will flow through each microvessel, and these few events cannot be missed if one wants to recover the whole vasculature. Our aberration correction method should thus increase the effective resolution of the images by localizing more bubbles in smaller vessels. As ULM is a long process, the shorter acquisition time enabled by the higher number of bubbles detected would lead to added comfort for the patient and the clinician.

To make the technique even more interesting from a clinical perspective, it should be implemented in real-time to improve the Ultrafast Doppler image. Empirically, we determined that 50 bubbles on average should be used to converge towards the correct phase aberration law, which gives a calculation time of ~60s for the total aberration correction process (on a computer with Intel® Xeon® CPU E5-2630 v3 @2.4 GHz). Once the aberration law is determined, however, it can be included in a classical real-time beamforming process to improve the Doppler image. The technique could thus be included in clinical practice by acquiring a few frames in the position of interest to retrieve the aberration correction parameters, after which the corrected images could be displayed in real-time for the clinician.

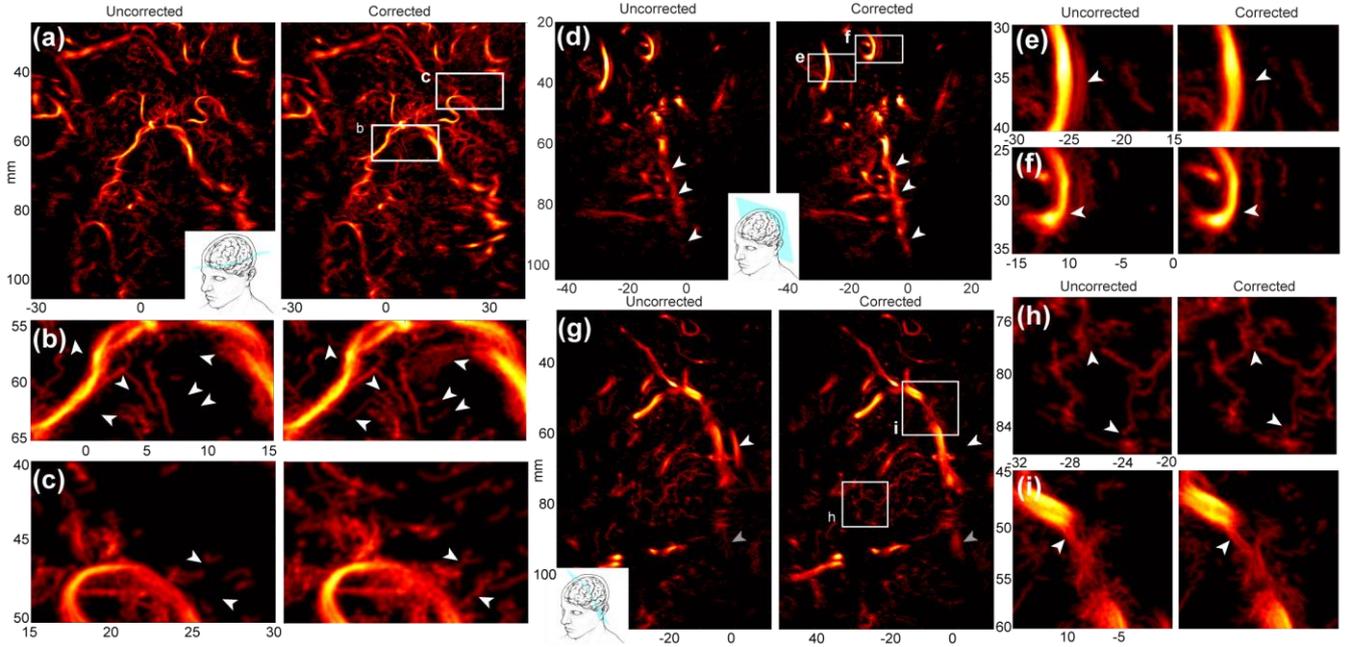

Fig. 7. Examples of microbubble localization images obtained without (left) and with (right) aberration correction, illustrating various improvements brought by the method. (a) Trans-temporal tilted axial view, with location of thumbnails b and c. (b) Some penetrating vessels of the mesencephalon invisible without aberration correction are revealed after correction (arrowheads). (c) Branching vessels presenting incomplete delineation without correction are improved after correction (arrowheads). (d) Posterior trans-temporal coronal view, with location of thumbnails e and f. Vessels faintly visible are strengthened after correction (arrowheads). (e) "Ghost" vessel is removed after aberration correction (arrowhead). (f) Similarly, some blurred area become sharper after aberration correction. (g) Trans-temporal tilted coronal view, with location of thumbnails h and i. White arrowhead point to another example of ghost vessel, removed after aberration correction, while grey arrowhead point to a reinforced slightly out of plane vessel. (h) Aberration correction recovers more complete (arrowheads) vessel paths, giving improved functional information about vascular structures. (i) In intermediate vessel sizes, structures appear less fuzzy after correction, with better delineation of the vascular bed. Acquisitions correspond to acquisitions 8 (a-b-c), 10 (d-e-f) and 7 (g-h-i) in Table 1. Tracks longer than 10 frames are represented.

To improve the images further, our method also derives an average speed of sound for each isoplanatic patch. This sound speed accounts for wave propagation in the skull bone and brain tissues and was always found in the $1520 - 1560\ m.s^{-1}$ range (for comparison, sound speed in the brain is around $1520\ m.s^{-1}$, and in the skull around $2700 - 3000\ m.s^{-1}$). This additional information is useful to provide a better image; however, it is not representative of each tissue property. In future work, it would be interesting to refine this sound speed calculation and provide sound speed maps of the different brain regions if possible. A possibility would be to compare the wavefronts obtained at different depths after phase aberration correction, to incrementally calculate the sound speed in tissue layers of different depths. As the sound speed is inherently linked to the tissue composition, and in particular to its fat fraction [38]], this quantitative information would give us further insight into brain tissues, with potential new diagnostic possibilities.

### 5) Conclusion

We proposed an adaptive aberration correction technique for skull bone aberrations in the adult human brain based on the backscattered signals coming from intravenously injected microbubbles. Our aberration correction technique improved image quality both for ultrafast Doppler imaging and Ultrasound Localization Microscopy, especially in cases with poor bone windows. For ULM, a 38% increase of long track microbubbles detected after aberration correction was found and allows an improved recording of brain hemodynamics in very small vessels. This technique is thus promising for better diagnosis and follow-up of brain pathologies such as aneurysms or stroke and could make transcranial ultrasound imaging possible even in patients particularly difficult to image.

**Acknowledgment**

The authors thank Hicham Serroune from Physics for Medicine Paris, for his help on probe holders and mechanical conception. The authors warmly thank the NVIDIA Corporation for their support through the NVidia GPU Grant program and the donation of a Titan Xp. This work was supported by the Fond National Suisse and the European Research Council (ERC) under the European Union's Seventh Framework Program (FP7/2007–2013)/ERC Advanced grant agreement no. 339244-FUSIMAGINE.

**References**

[1] D. S. Dede *et al.*, "Assessment of Endothelial Function in Alzheimer's Disease: Is Alzheimer's Disease a Vascular Disease?," *J. Am. Geriatr. Soc.*, vol. 55, no. 10, pp. 1613–1617, Oct. 2007, doi: 10.1111/j.1532-5415.2007.01378.x.

[2] B. Ding *et al.*, "Pattern of cerebral hyperperfusion in Alzheimer's disease and amnestic mild cognitive impairment using voxel-based analysis of 3D arterial spin-labeling imaging: initial experience," *Clin. Interv. Aging*, vol. 9, p. 493, Mar. 2014, doi: 10.2147/CIA.S58879.

[3] X. Zhu *et al.*, "Vascular oxidative stress in Alzheimer disease.," *J. Neurol. Sci.*, vol. 257, no. 1–2, pp. 240–6, Jun. 2007, doi: 10.1016/j.jns.2007.01.039.

[4] J. Guan *et al.*, "High-resolution magnetic resonance imaging of intracranial aneurysms treated by flow diversion," *Interdiscip. Neurosurg.*, vol. 10, pp. 69–74, Dec. 2017, doi: 10.1016/J.INAT.2017.07.004.

[5] A. P. Tregaskiss, A. N. Goodwin, L. D. Bright, C. H. Ziegler, and R. D. Acland, "Three-dimensional CT angiography: A new technique for imaging microvascular anatomy," *Clin. Anat.*, vol. 20, no. 2, pp. 116–123, Mar. 2007, doi: 10.1002/ca.20350.

[6] A. D'Andrea *et al.*, "Transcranial Doppler ultrasonography: From methodology to major clinical applications," *World J. Cardiol.*, vol. 8, no. 7, p. 383, 2016, doi: 10.4330/wjc.v8.i7.383.

[7] T. L. Szabo, "Diagnostic Ultrasound Imaging: Inside Out Second Edition," 2014. [Online]. Available: www.elsevierdirect.com/rights.

[8] J. Bercoff *et al.*, "Ultrafast compound Doppler imaging: providing full blood flow characterization," *IEEE Trans. Ultrason. Ferroelectr. Freq. Control*, vol. 58, no. 1, pp. 134–147, Jan. 2011, doi: 10.1109/TUFFC.2011.1780.

[9] A. C. H. Yu and L. Lovstakken, "Eigen-based clutter filter design for ultrasound color flow imaging: A review," *IEEE Trans. Ultrason. Ferroelectr. Freq. Control*, vol. 57, no. 5, pp. 1096–1111, 2010, doi: 10.1109/TUFFC.2010.1521.

[10] J. Baranger, B. Arnal, F. Perren, O. Baud, M. Tanter, and C. Demene, "Adaptive Spatiotemporal SVD Clutter Filtering for Ultrafast Doppler Imaging Using Similarity of Spatial Singular Vectors," *IEEE Trans. Med. Imaging*, vol. 37, no. 7, pp. 1574–1586, 2018, doi: 10.1109/TMI.2018.2789499.

[11] C. Demené, J. Mairesse, J. Baranger, M. Tanter, and O. Baud, "Ultrafast Doppler for neonatal brain imaging," *NeuroImage*. 2019, doi: 10.1016/j.neuroimage.2018.04.016.

[12] O. Couture, B. Besson, G. Montaldo, M. Fink, and M. Tanter, "Microbubble ultrasound super-localization imaging (MUSLI)," *IEEE Int. Ultrason. Symp. IUS*, pp. 1285–1287, 2011, doi: 10.1109/ULTSYM.2011.6293576.

[13] Y. Desailly, O. Couture, M. Fink, and M. Tanter, "Sono-activated ultrasound localization microscopy," *Appl. Phys. Lett.*, vol. 103, no. 17, p. 174107, Oct. 2013, doi: 10.1063/1.4826597.

[14] C. Errico *et al.*, "Ultrafast ultrasound localization microscopy for deep super-resolution vascular imaging," 2015, doi: 10.1038/nature16066.

[15] C. Demené *et al.*, "Transcranial ultrafast ultrasound localization microscopy of brain vasculature in patients," *Nat. Biomed. Eng.*, vol. 5, no. 3, pp. 219–228, 2021, doi: 10.1038/s41551-021-00697-x.

[16] E. Betzig *et al.*, "Imaging Intracellular Fluorescent Proteins at Nanometer Resolution\r10.1126/science.1127344," *Science (80-. ).*, vol. 313, no. 5793, pp. 1642–1645, 2006, [Online]. Available: d:%5CUsers%5CAndras%5CArticles%5CScience%5C2006%5CScience 313 1642 (2006) Betzig.pdf.

[17] M. A. O'Reilly and K. Hynynen, "A super-resolution ultrasound method for brain vascular mapping.," *Med. Phys.*, vol. 40, no. 11, p. 110701, Nov. 2013, doi: 10.1118/1.4823762.

[18] O. M. Viessmann, R. J. Eckersley, K. Christensen-Jeffries, M. X. Tang, and C. Dunsby, "Acoustic super-resolution with ultrasound and microbubbles," *Phys. Med. Biol.*, vol. 58, no. 18, pp. 6447–6458, Sep. 2013, doi: 10.1088/0031-9155/58/18/6447.

[19] S. W. Flax and M. O'Donnell, "Phase-aberration correction using signals from point reflectors and diffuse scatterers: Basic principles," *Ultrason. Ferroelectr. Freq. Control. IEEE Trans.*, vol. 35, no. 6, pp. 758–767, 1988, doi: 10.1109/58.9333.

[20] G. C. C. Ng, S. S. S. Worrell, P. D. D. Freiburger, and G. E. E. Trahey, "A comparative evaluation of several algorithms for phase aberration correction," *IEEE Trans. Ultrason. Ferroelectr. Freq. Control*, vol. 41, no. 5, pp. 631–643, 1994, doi: 10.1109/58.308498.

[21] R. Mallart and M. Fink, "Adaptive focusing in scattering media through sound-speed inhomogeneities: The van Cittert Zernike approach and focusing criterion," *J. Acoust. Soc. Am.*, vol. 96, no. 6, pp. 3721–3732, 1994, doi: 10.1121/1.410562.

[22] L. Nock, G. E. Trahey, and S. W. Smith, "Phase aberration correction in medical ultrasound using speckle brightness as a quality factor," *J. Acoust. Soc. Am.*, vol. 85, no. 5, pp. 1819–1833, May 1989, doi: 10.1121/1.397889.

[23] J. R. Sukovich, Z. Xu, T. L. Hall, J. J. Macoskey, and C. A. Cain, "Transcranial histotripsy acoustic-backscatter localization and aberration correction for volume treatments," *J. Acoust. Soc. Am.*, vol. 141, no. 5, pp. 3490–3490, May 2017, doi: 10.1121/1.4987285.

[24] M. Pernot, G. Montaldo, M. Tanter, and M. Fink, "Ultrasonic stars for time reversal focusing using induced cavitation bubbles," *AIP Conf. Proc.*, vol. 829, no. 223, 2006, doi: 10.1063/1.2205470.

[25] B. F. Osmanski, G. Montaldo, M. Tanter, and M. Fink, "Aberration correction by time reversal of moving speckle noise," *IEEE Trans. Ultrason. Ferroelectr. Freq. Control*, vol. 59, no. 7, pp. 1575–1583, 2012, doi: 10.1109/TUFFC.2012.2357.

[26] by Nikolas Moravek Ivancevich, S. W. Smith, G. E.

Trahey, K. R. Nightingale, D. T. Laskowitz, and S. Mukundan, "Phase Aberration Correction for Real-Time 3D Transcranial Ultrasound Imaging," 2009.

[27] N. M. Ivancevich, G. F. Pinton, H. A. Nicoletto, E. Bennett, D. T. Laskowitz, and S. W. Smith, "Real-Time 3-D Contrast-Enhanced Transcranial Ultrasound and Aberration Correction," *Ultrasound Med. Biol.*, vol. 34, no. 9, pp. 1387–1395, 2008, doi: 10.1016/j.ultrasmedbio.2008.01.015.

[28] B. D. Lindsey, H. A. Nicoletto, E. R. Bennett, D. T. Laskowitz, and S. W. Smith, "3-D Transcranial Ultrasound Imaging with Bilateral Phase Aberration Correction of Multiple Isoplanatic Patches: A Pilot Human Study with Microbubble Contrast Enhancement," *Ultrasound Med. Biol.*, vol. 40, no. 1, pp. 90–101, Jan. 2014, doi: 10.1016/j.ultrasmedbio.2013.09.006.

[29] C. Demene *et al.*, "Spatiotemporal clutter filtering of ultrafast ultrasound data highly increases Doppler and fUltrasound sensitivity," *IEEE Trans. Med. Imaging*, vol. PP, no. 99, pp. 1–1, 2015, doi: 10.1109/TMI.2015.2428634.

[30] T. Jerman, F. Pernus, B. Likar, and Z. Spiclin, "Enhancement of Vascular Structures in 3D and 2D Angiographic Images," *IEEE Trans. Med. Imaging*, vol. 35, no. 9, pp. 2107–2118, Sep. 2016, doi: 10.1109/TMI.2016.2550102.

[31] B. Heiles *et al.*, "Ultrafast 3D Ultrasound Localization Microscopy Using a 32x32 Matrix Array," *IEEE Trans. Med. Imaging*, vol. 38, no. 9, pp. 2005–2015, 2019, doi: 10.1109/tmi.2018.2890358.

[32] R. Mallart and M. Fink, "The van Cittert–Zernike theorem in pulse echo measurements," *J. Acoust. Soc. Am. J. Acoust. Soc. Am. Acoust. Soc. Am. Acoust. Soc. Am.*, vol. 90, no. 101, pp. 2718–1847, 1991, doi: 10.1121/1.418235.

[33] A. Derode and M. Fink, "Spatial coherence of ultrasonic speckle in composites," *IEEE Trans. Ultrason. Ferroelectr. Freq. Control*, vol. 40, no. 6, pp. 666–675, Nov. 1993, doi: 10.1109/58.248209.

[34] "Robust sound speed estimation for ultrasound-based hepatic steatosis assessment."

[35] Y. Desailly, J. Pierre, O. Couture, and M. Tanter, "Resolution limits of ultrafast ultrasound localization microscopy," *Phys. Med. Biol.*, vol. 60, no. 22, pp. 8723–8740, Oct. 2015, doi: 10.1088/0031-9155/60/22/8723.

[36] A. Duck, Francis A., *Physical Properties of Tissues: A Comprehensive Reference Network*. 1990.

[37] V. Hingot, C. Errico, B. Heiles, L. Rahal, M. Tanter, and O. Couture, "Microvascular flow dictates the compromise between spatial resolution and acquisition time in Ultrasound Localization Microscopy," *Sci. Rep.*, 2019, doi: 10.1038/s41598-018-38349-x.

[38] M. D. B. Marion Imbault1,7 *et al.*, "Ultrasonic fat fraction quantification using in vivo adaptive sound speed estimation," *Phys. Med. Biol.*, 2018, doi: 10.1088/1361-6560/aae661.

**Supplemental Figures**

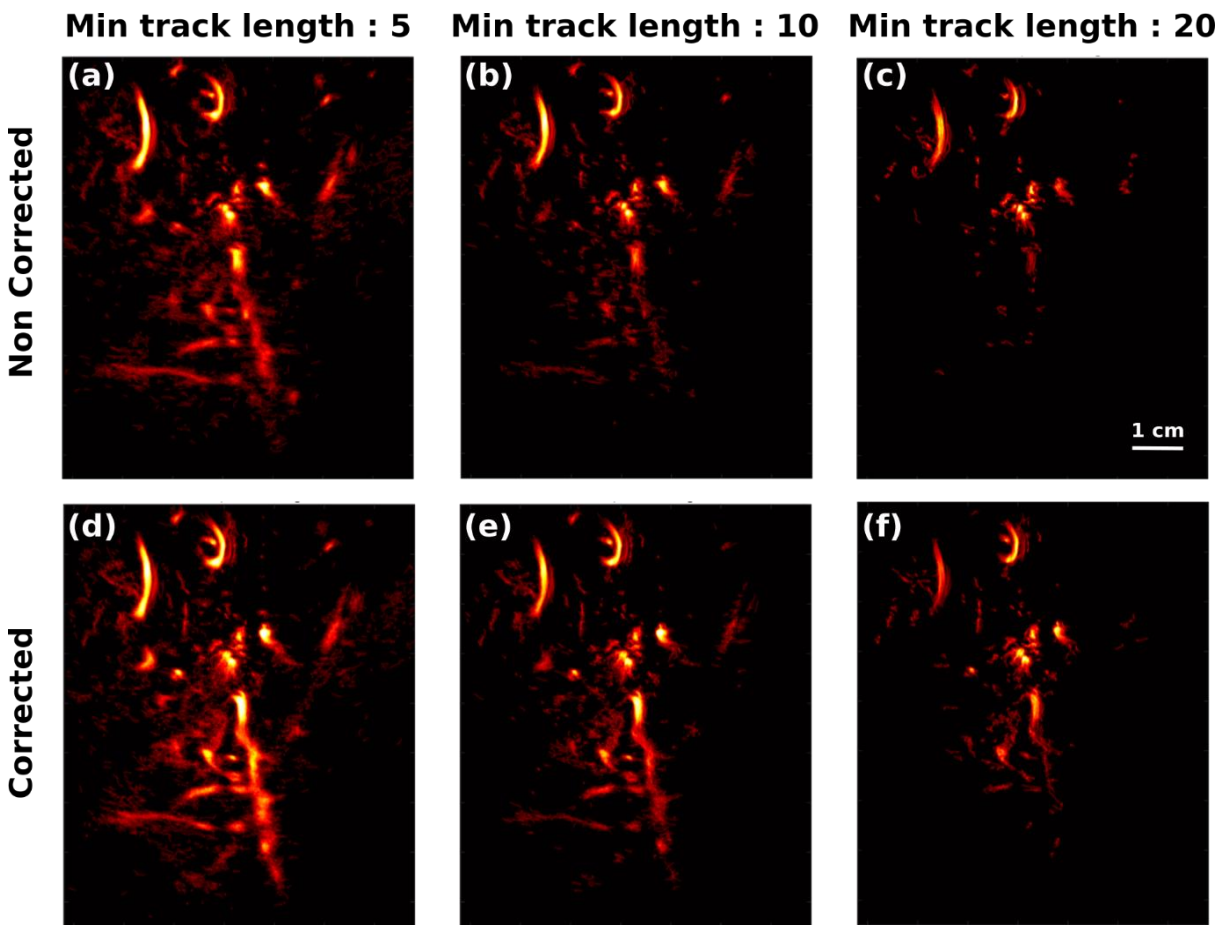

Supplemental Figure 1: Microbubble localization images obtained for increasing values of minimum track lengths. (a), (b) and (c) respectively show the non-corrected images with minimum track lengths of 5, 10 and 20 frames. (d), (e) and (f) respectively show the corrected images with minimum track lengths of 5, 10 and 20 frames.

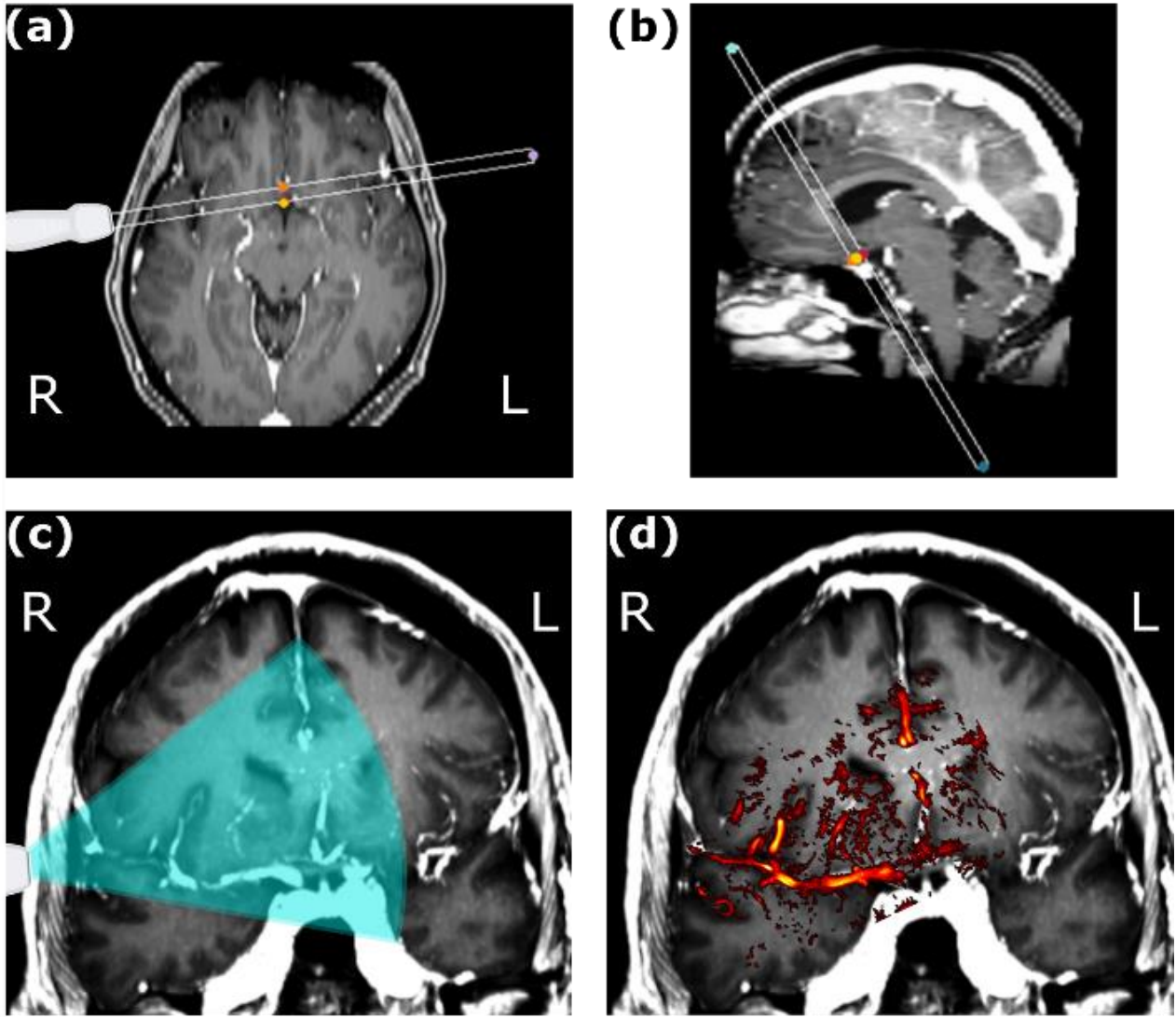

Supplemental Figure 1: MRA images for the patient in Figure 7g to 7i. (a) and (b) respectively show axial and sagittal slices, where the ultrasonic imaging plane is represented as a slab. (c) and (d) show maximum intensity projection of the slab depicted in (a) and (b). Ultrasonic field of view is represented as a light blue overlay on the left image, and the vasculature extracted from the localization images in Fig. 7 – g is overlaid on the right image.

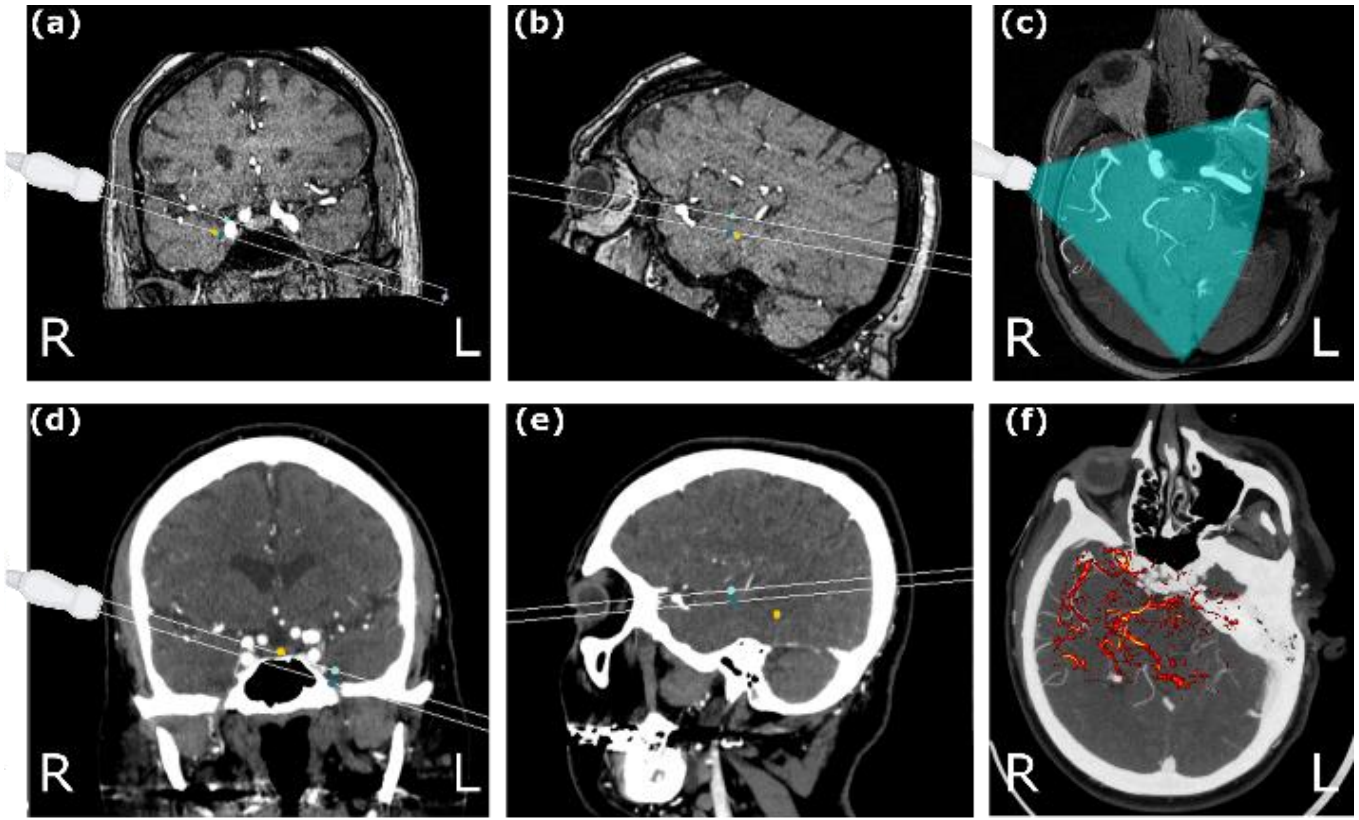

Supplemental Figure 2: Top: MRA images for the patient in Figure 7a to 7f. (a) and (b) respectively show coronal and sagittal slices, where the ultrasonic imaging plane is represented as a slab. (c) shows the maximum intensity projection of the slab depicted in (a) and (b). Ultrasonic field of view is represented as a light blue overlay. Bottom: CT images of the same patient. (d) and (e) respectively show coronal and sagittal slices, where the ultrasonic imaging plane is represented as a slab. (f) shows the maximum intensity projection of the slab depicted in (d) and (f) and the vasculature extracted from the localization images in Fig. 7 – a is overlaid